\documentclass[]{spie}  

\usepackage{amsmath,amsfonts,amssymb}
\usepackage{graphicx}
\usepackage[colorlinks=true, allcolors=blue]{hyperref}

\usepackage{float}
\usepackage{subcaption}

\title{Score-Based Ideal Observer Approximation 
via Denoising Score Matching
for Signal-Known-Exactly Detection Tasks}

\author[a, b]{Weimin Zhou}
\affil[a]{Wyant College of Optical Sciences, University of Arizona, AZ 85721, USA}
\affil[b]{Department of Radiology and Imaging Sciences, University of Arizona College of Medicine -- Tucson, AZ 85721, USA}

\authorinfo{Further author information: (Send correspondence to Weimin Zhou)\\Weimin Zhou: E-mail: weiminzhou@arizona.edu}

\begin{document} 
\maketitle

\begin{abstract}
The Bayesian Ideal Observer (IO) establishes the theoretical upper bound on task performance for binary detection tasks. 
However, analytical computation of the IO test statistic is generally intractable. Numerical approaches based on Markov-chain Monte Carlo (MCMC) methods, including their recent deep generative model-based extensions, typically require extensive posterior sampling for each test image. Supervised learning has also been investigated to approximate the IO performance. However, such methods are typically trained for a specific detection task and signal and may require retraining when the task or signal changes.
The score function, defined as the gradient of the log probability density, encodes the local geometry of the data distribution and is a fundamental quantity in modern score-based generative modeling.
This work reformulates the IO test statistic in terms of the score function and introduces a score-based ideal observer (SIO). The proposed SIO uses a denoising convolutional neural network trained exclusively on signal-absent images to estimate the signal-absent score function. 
Once trained, the resulting score model can be used to approximate the IO test statistic for detection tasks involving arbitrary additive signals, without per-image posterior sampling or signal-specific retraining. Numerical studies consider a signal-known-exactly (SKE) detection task with a stochastic lumpy-background model. The results demonstrate that the proposed SIO can closely approximate the IO performance. 
\end{abstract}


\section{INTRODUCTION}
\label{sec:intro}  
The Bayesian ideal observer (IO) provides the optimal decision strategy for binary signal detection tasks and establishes a theoretical upper bound on task performance for assessing and optimizing imaging systems \cite{barrett2013foundations}. However, analytical computation of the IO test statistic is generally intractable for imaging problems involving complex stochastic backgrounds because the underlying probability density functions are typically unknown.
Existing numerical approaches based on Markov-chain Monte Carlo (MCMC) \cite{kupinski2003ideal, zhou2020markov, li2025approximating} can approximate the IO test statistic but typically require extensive posterior sampling for each test image. Supervised learning methods can provide efficient alternatives \cite{kupinski2001ideal, zhou2019approximating, zhou2020approximating}, but are typically trained for a specific detection task and signal and may require retraining when the task or signal changes.

Recent advances in score-based generative modeling provide an alternative means of characterizing complex data distributions through the score function, defined as the gradient of the log probability density \cite{song2019generative, song2020score}. The score function can be learned directly from data using denoising score matching without requiring explicit knowledge of the underlying probability density \cite{vincent2011connection}. 

In this work, we propose a score-based formulation of the IO test statistic and develop a score-based ideal observer (SIO) for approximating the IO. The proposed formulation expresses the IO test statistic using the signal-absent score function integrated along a path defined by the known signal, followed by a nonprewhitening matched filter (NPWMF) operation.  A denoising convolutional neural network is trained exclusively on signal-absent images to estimate the score function. Once trained, the same score model can be used to approximate the IO for detection tasks involving arbitrary additive signals without task-specific supervised retraining.

\section{Methods}  

\subsection{Score and Denoising Score Matching}
\label{sec:title}

Recent advances in generative modeling have extensively utilized the score function, defined as the gradient of the log probability density with respect to the image data. The score function characterizes how the probability density of an image changes with respect to local perturbations in the image space and therefore serves as a powerful representation of underlying image statistics. For image data $\mathbf{x}$ sampled from the probability density function $\text{pr}(\mathbf{x})$, the score function is defined as:
\begin{equation}
\psi_{\mathbf{x}}(\mathbf{x}) = \nabla_\mathbf{x} \log \text{pr}(\mathbf{x}).
\end{equation}
To approximate the score function $\psi(\mathbf{x})$, a network $\phi_\theta: \mathbb{R}^M \rightarrow \mathbb{R}^M$ parameterized by $\theta$ can be trained via score matching. 
However, directly optimizing the original score-matching objective requires computing the trace of the Jacobian of the network, which is not scalable to large networks and high-dimensional data \cite{song2019generative}. Denoising score matching (DSM) avoids this difficulty by introducing a known perturbation distribution  $q_\sigma (\tilde{\mathbf{x}} | \mathbf{x})$ to produce the perturbed data $\tilde{\mathbf{x}}$ from $\mathbf{x}$, and 
the network parameters are optimized by minimizing \cite{song2019generative}:
\begin{equation}
\label{eq:denoising-score-matching}
\mathbb{E}_{q_\sigma(\tilde{\mathbf{x}}\mid\mathbf{x})\text{pr}(\mathbf{x})}
\Big[ \big\|{\phi}_\theta(\tilde{\mathbf{x}})-\nabla_{\tilde{\mathbf{x}}}\log q_\sigma(\tilde{\mathbf{x}}\mid\mathbf{x})\big\|_2^2 \Big].
\end{equation}
A common choice for the perturbation distribution $q_\sigma (\tilde{\mathbf{x}} | \mathbf{x})$ is an isotropic Gaussian distribution with standard deviation $\sigma$. The corresponding conditional score is:
\begin{equation}
\nabla_{\tilde{\mathbf{x}}}
\log q_\sigma(\tilde{\mathbf{x}}\mid\mathbf{x})
=
-\frac{\tilde{\mathbf{x}}-\mathbf{x}}{\sigma^2}.
\end{equation}
The denoising score matching objective in Eq. (\ref{eq:denoising-score-matching}) becomes a least-squares loss for predicting the scaled noise.

Below, we first reformulate the IO test statistic for signal-known-exactly (SKE) detection tasks using the score function and then present an IO approximation approach using denoising score matching.

\subsection{IO Approximation via Denoising Score Matching}
Consider an SKE signal detection task in which an observer classifies imaging data $\mathbf{g}$ as arising from either the signal-absent hypothesis ($H_0$) or the signal-present hypothesis ($H_1$):
\begin{equation}\label{eq:cpt2_binary}
\begin{split}
&H_{0}:  \mathbf{g} =  \mathbf{H}\mathbf{f}_b+ \mathbf{n} \equiv \mathbf{b} + \mathbf{n} \equiv \mathbf{f}_0 + \mathbf{n}, \\
&H_{1}:  \mathbf{g} =  \mathbf{H}(\mathbf{f}_b + \mathbf{f}_s)+ \mathbf{n} \equiv \mathbf{b} + \mathbf{s} + \mathbf{n} \equiv \mathbf{f}_1 + \mathbf{n}. 
\end{split}
\end{equation}
Assume that the measurement $\mathbf g$ is continuously valued, and denote its probability density under $H_j$ by $\text{pr}(\mathbf{g} | H_j)$, ($j=0, 1$).
For a SKE detection task in which the signal $\mathbf{s}$ is deterministic and additive, the signal-present distribution is a translated version of the signal-absent distribution:
\begin{equation}
\text{pr}(\mathbf{g} | H_1) =  
\text{pr}_{\mathbf{f}_1 + \mathbf{n}}(\mathbf{g} ) =
\text{pr}_{\mathbf{f}_0 + \mathbf{s} + \mathbf{n}}(\mathbf{g} ) 
=
\text{pr}_{\mathbf{f}_0  + \mathbf{n}}(\mathbf{g} - \mathbf{s}) =
\text{pr}(\mathbf{g} - \mathbf{s}| H_0)
\end{equation} 
The IO test statistic, defined by the log-likelihood ratio, can then be expressed as:
\begin{equation}
\lambda_\text{IO}(\mathbf{g}) = \log\left[\frac{\text{pr}(\mathbf{g}|H_1)}{\text{pr}(\mathbf{g}|H_0)}\right] = 
\log\left[\frac{\text{pr}(\mathbf{g} - \mathbf{s}| H_0)}{\text{pr}(\mathbf{g}| H_0)}\right ] = \log\left[\text{pr}(\mathbf{g} - \mathbf{s}| H_0) \right]- \log\left[\text{pr}(\mathbf{g}| H_0) \right].
\end{equation}
Let $\mathbf{g}_s(\alpha) = \mathbf{g} - \alpha \mathbf{s}, \ \ \alpha \in [0,1]$ denote the straight-line path connecting $\mathbf{g}$ and $\mathbf{g} - \mathbf{s}$. Then, by the fundamental theorem of calculus,
\begin{align}
\label{eq:sio}
\lambda_\text{IO}(\mathbf{g}) 
&= 
\log \!\left[\operatorname{pr}\!\left(\mathbf g_s(1)\mid H_0\right)\right]
-
\log \!\left[\operatorname{pr}\!\left(\mathbf g_s(0)\mid H_0\right)\right] 
\nonumber\\
&=
\int_{0}^{1}
\frac{d}{d\alpha}
\log \!\left[
\operatorname{pr}\!\left(\mathbf g_s(\alpha)\mid H_0\right)
\right]
\, d\alpha
\nonumber\\
&=
\int_{0}^{1}
\nabla_{\mathbf g_s(\alpha)}
\log \!\left[
\operatorname{pr}\!\left(\mathbf g_s(\alpha)\mid H_0\right)
\right]^T
\frac{d\mathbf g_s(\alpha)}{d\alpha}
\, d\alpha 
\nonumber\\
&=
-
\mathbf s ^T
\int_{0}^{1}
\boldsymbol{\psi}_{\mathbf g\mid H_0}
\!\left(\mathbf g_s(\alpha)\right) d\alpha
\,
\end{align}
where $\boldsymbol{\psi}_{\mathbf g\mid H_0}\left(\mathbf g_s(\alpha)\right) \triangleq \nabla_{\mathbf g_s(\alpha)}
\log \!\left[
\operatorname{pr}\left(\mathbf g_s(\alpha)\mid H_0\right)
\right]$ denotes the score function of the signal-absent measurement distribution evaluated at $\mathbf g_s(\alpha)$.
Equation (\ref{eq:sio}) shows that the IO test statistic can be computed by integrating the signal-absent score function along the path connecting $\mathbf{g}$ and $\mathbf{g} - \mathbf{s}$, followed by an inner product with the negative signal image.
The straight-line path is used here. In principle, any continuously differentiable path connecting $\mathbf{g}$ and $\mathbf{g} - \mathbf{s}$ may also be used.

\begin{figure}[H]
     \centering
         \centering
         \includegraphics[width=0.95\textwidth]{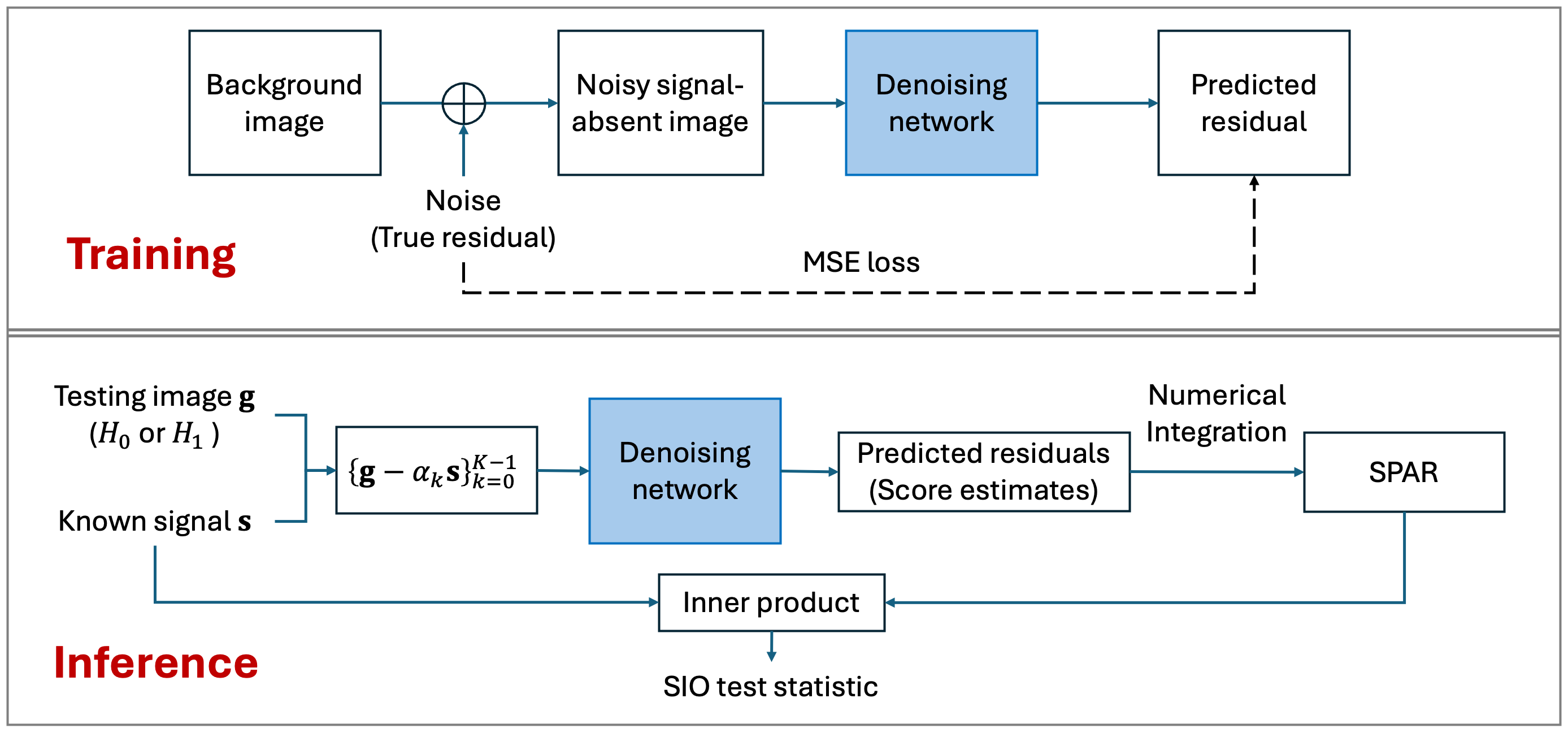}
         \vspace{0.2cm}
         \caption{ 
         Overview of the proposed score-based ideal observer (SIO) framework. During training, a denoising convolutional neural network is trained using noisy signal-absent images, where the corresponding noise realizations serve as the training targets. During inference, the trained network is evaluated at multiple signal-interpolated images to estimate the corresponding residuals (proportional to the negative score function). The predicted residuals are numerically integrated to produce the signal-path averaged residual (SPAR), which is subsequently combined with the known signal through an inner product to compute the SIO test statistic.}
         \label{fig:sio}
     \end{figure}

To estimate the score function $\boldsymbol{\psi}_{\mathbf g\mid H_0}\left(\mathbf g\right)$, we regard a signal-absent measurement $\mathbf{g}$ as a perturbed realization of the corresponding noise-free image $\mathbf{f}_0$. 
This is possible when evaluating imaging systems through virtual imaging trials, in which samples of $\mathbf{f}_{0}$ can be generated using specified object and imaging models, and the corresponding measurements $\mathbf{g}$ can then be simulated by adding measurement noise drawn from a known distribution.
Given a set of paired samples
$\{(\mathbf f_0^{(i)},\mathbf g^{(i)})\}_{i=1}^{N}$ drawn from $\text{pr}(\mathbf{g}|\mathbf{f}_0, H_0)\text{pr}(\mathbf{f}_0)$, the DSM formulation can be applied by identifying
$\mathbf f_0$ as the unperturbed data and $\mathbf g$ as the perturbed
data. Accordingly, the score network
$\boldsymbol{\phi}_{\boldsymbol{\theta}}(\mathbf g)$ is trained by
minimizing
\begin{equation}
\widehat{\mathcal L}_{\mathrm{DSM}}(\boldsymbol{\theta})
=
\frac{1}{N}
\sum_{i=1}^{N}
\left\|
\phi_{\boldsymbol{\theta}}
\left(\mathbf g^{(i)}\right)
-
\nabla_{\mathbf g^{(i)}}
\log
\operatorname{pr}
\left(
\mathbf g^{(i)}
\mid
\mathbf f_0^{(i)}, H_0
\right)
\right\|_2^2 .
\label{eq:dsm_io_objective}
\end{equation}
The population minimizer
of this objective estimates the marginal score of the signal-absent measurement distribution. Once the model has been successfully trained, we have
\begin{equation}
\boldsymbol{\phi}_{\boldsymbol{\theta}}(\mathbf g)
\approx
\nabla_{\mathbf g}
\log
\operatorname{pr}(\mathbf g\mid H_0)
=
\boldsymbol{\psi}_{\mathbf g\mid H_0}(\mathbf g).
\label{eq:estimated_signal_absent_score}
\end{equation}
When the measurement noise components are independent and identically distributed (i.i.d.) Gaussian random variables with zero mean and variance \(\sigma_n^2\), the conditional score is
\begin{equation}
\nabla_{\mathbf g^{(i)}}
\log
\operatorname{pr}
\left(
\mathbf g^{(i)}
\mid
\mathbf f_0^{(i)}, H_0
\right) = 
-\frac{
\mathbf g^{(i)}-\mathbf f_0^{(i)}
}{
\sigma_n^2
}
= -\frac{1}{\sigma_n^2}\mathbf{n}^{(i)}.
\end{equation}
In this case, we can train the network using the measurement residual. Specifically, let $\mathbf r_{\boldsymbol{\theta}}(\mathbf g)$ denote a neural network parameterized by weight parameters $\boldsymbol{\theta}$. The network is trained to estimate the measurement noise by minimizing:
\begin{equation}
\widehat{\mathcal L}_{r}(\boldsymbol{\theta})
=
\frac{1}{N}
\sum_{i=1}^{N}
\left\|
\mathbf r_{\boldsymbol{\theta}}
\left(\mathbf g^{(i)}\right)
-
\left(
\mathbf g^{(i)}-\mathbf f_0^{(i)}
\right)
\right\|_2^2.
\label{eq:residual_prediction_loss}
\end{equation}
After training, the score function can be approximated as: $\boldsymbol{\psi}_{\mathbf g\mid H_0}(\mathbf g) \approx -\frac{1}{\sigma_n^2} \mathbf{r}_{\boldsymbol{\theta}} (\mathbf{g})$. 
Using the trained residual network, the IO test statistic in Eq. (\ref{eq:sio}) can be approximated by numerical integration. In this work, the left Riemann sum is employed. Let $K$ denote the number of integration points. We then have:
\begin{equation}
\widehat{\lambda}_{\mathrm{IO}}(\mathbf g) = \frac{1}{\sigma_n^2} \mathbf{s}^T  \Big[\frac{1}{K}\sum_{k=0}^{K-1} \mathbf{r}_{\boldsymbol{\theta}}\left(\mathbf{g} -\alpha_k\mathbf{s}\right) \Big]
=
\frac{1}{\sigma_n^2}\mathbf{s}^T\bar{\mathbf{r}}_{\boldsymbol{\theta},K}(\mathbf{g};\mathbf{s}), 
\label{eq:score_based_io}
\end{equation}
where $\alpha_k = \frac{k}{K}$ and  $\bar{\mathbf{r}}_{\boldsymbol{\theta},K}(\mathbf{g};\mathbf{s}) \triangleq \frac{1}{K}\sum_{k=0}^{K-1} \mathbf{r}_{\boldsymbol{\theta}}\left(\mathbf{g} -\alpha_k\mathbf{s}\right)$ is referred to as the signal-path-averaged residual (SPAR).
Remarkably, Eq. (\ref{eq:score_based_io}) reveals an elegant interpretation of the IO test statistic: once the SPAR has been estimated, the complex IO computation reduces to a simple non-prewhitening matched-filter operation applied to the SPAR. The overview of the proposed SIO framework is shown in Fig. \ref{fig:sio}.

\section{Numerical studies and results}
\label{sec:sections}
Computer-simulation studies were conducted to evaluate the proposed score-based ideal observer (SIO) approximation. A signal-known-exactly (SKE) binary signal detection task with a stochastic lumpy background \cite{rolland1992effect} was considered. A residual denoising convolutional neural network was trained exclusively on signal-absent images to estimate the signal-absent score function. The trained network was subsequently used to approximate the IO test statistic using the left Riemann-sum approximation of the proposed line-integral formulation. Observer performance was quantified by the area under the receiver operating characteristic curve (AUC). The simulation setup and corresponding results are described below.
 
\subsection{Simulation Setup}

\begin{figure}[H]
     \centering
         \centering
         \includegraphics[width=\textwidth]{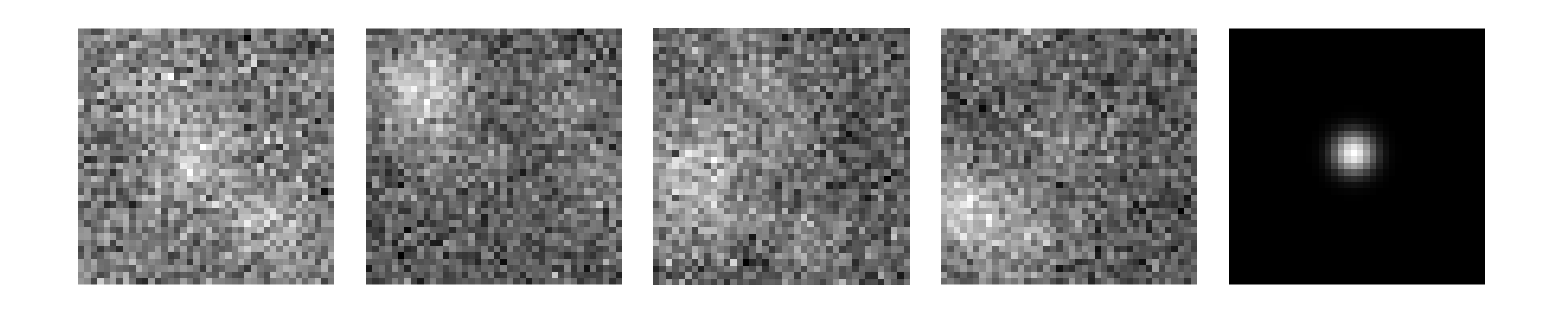}
         \caption{ 
         From left to right are four examples of signal-present noisy images and the signal image.}
         \label{fig:simulation}
     \end{figure}

A Type-I lumpy background model \cite{rolland1992effect} was employed to generate signal-absent object images. The background object was modeled as
$f_b(\mathbf{r})=\sum_{n=1}^{N_b} l(\mathbf{r}-\mathbf{r}_n)$, where $N_b$ is a Poisson random variable with mean $\bar{N}=5$, and $\mathbf{r}_n$ denotes the center location of the $n$th lump, sampled from a uniform distribution over a $40\times40$ field of view (FOV). Each lump was modeled by a 2D Gaussian function with an amplitude of $1.2$ and a width of $4.8$.
A signal-known-exactly (SKE) binary detection task was considered. The signal was modeled as a 2D Gaussian function centered in the field of view with an amplitude of 0.6 and a width of~2.0.
The imaging system was modeled as a continuous-to-discrete Gaussian blur operator, $
h_m(\mathbf{r})
=
\frac{h}{2\pi w^2}
\exp\!\left(
-\frac{(\mathbf{r}-\mathbf{r}_m)^T(\mathbf{r}-\mathbf{r}_m)}
{2w^2}
\right),
$
where $h=1.5$, $w=0.8$, and $\mathbf{r}_m$ denotes the spatial location of the $m$th image pixel that uniformly samples the FOV. The resulting measurement images consisted of $40\times40$ pixels. Independent Gaussian noise having zero mean and a standard deviation of $1.3$ was added to the noise-free images to produce the noisy image data. 
Examples of the signal-present images and the signal are shown in Fig.~\ref{fig:simulation}.

\subsection{Network training}
A 17-layer DnCNN \cite{zhang2017beyond} was employed to estimate the residual (noise) image required for score computation. The network architecture followed the original DnCNN design, consisting of an initial convolutional layer with ReLU activation, 15 convolutional layers each followed by batch normalization and ReLU activation, and a final convolutional layer for residual prediction. All convolutional layers used $3\times3$ kernels.
The DnCNN was trained using 100,000 independently generated background (signal-absent) images. During training, independent measurement noise was added to each background image on-the-fly to generate the noisy input, while the corresponding noise realization was used as the training target. The network parameters were optimized by minimizing the mean squared error (MSE) between the predicted and true residual images using the Adam optimizer with an initial learning rate of $10^{-3}$.
The training was performed on a single NVIDIA L40S GPU. The trained network was subsequently employed to approximate the IO test statistic using the proposed score-based formulation.

\subsection{Results}

\begin{figure}[H]
     \centering
     \begin{subfigure}[b]{0.48\textwidth}
         \centering
         \includegraphics[width=\textwidth]{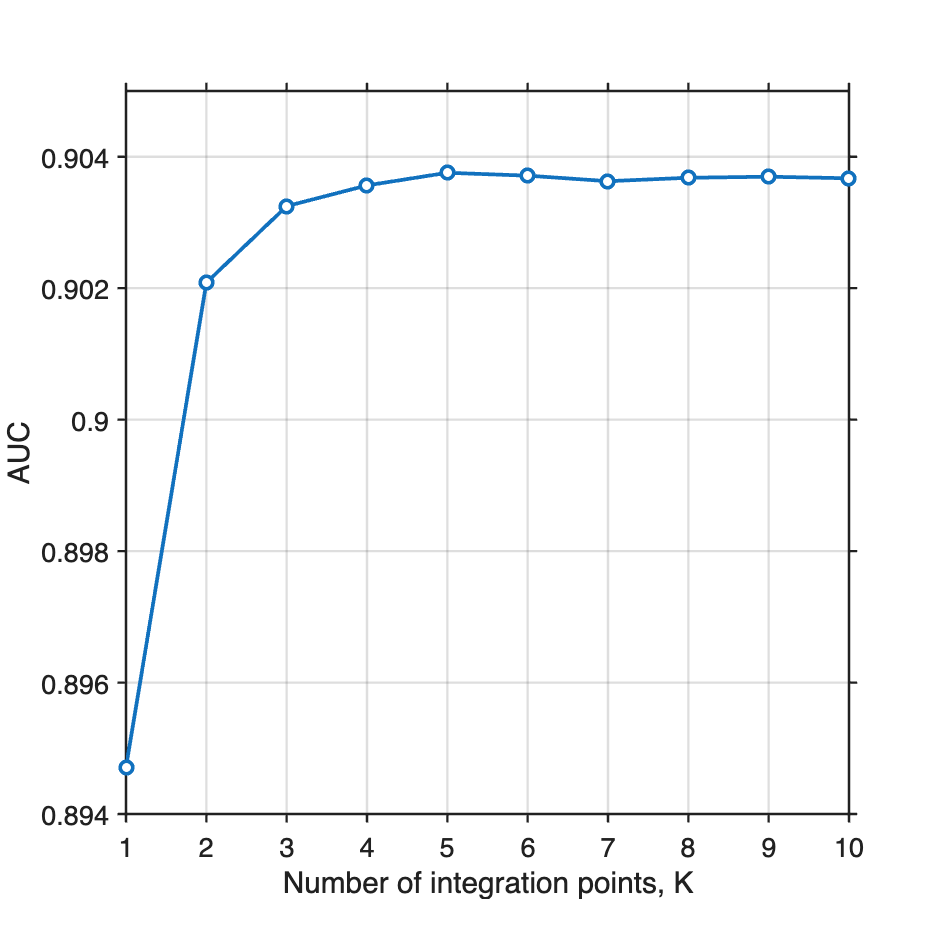}
         \vspace{-0.8cm}
         \caption{AUC versus the number of integration points $K$.}
     \end{subfigure}
     \hspace{0.02\textwidth}
     \begin{subfigure}[b]{0.48\textwidth}
         \centering
         \includegraphics[width=\textwidth]{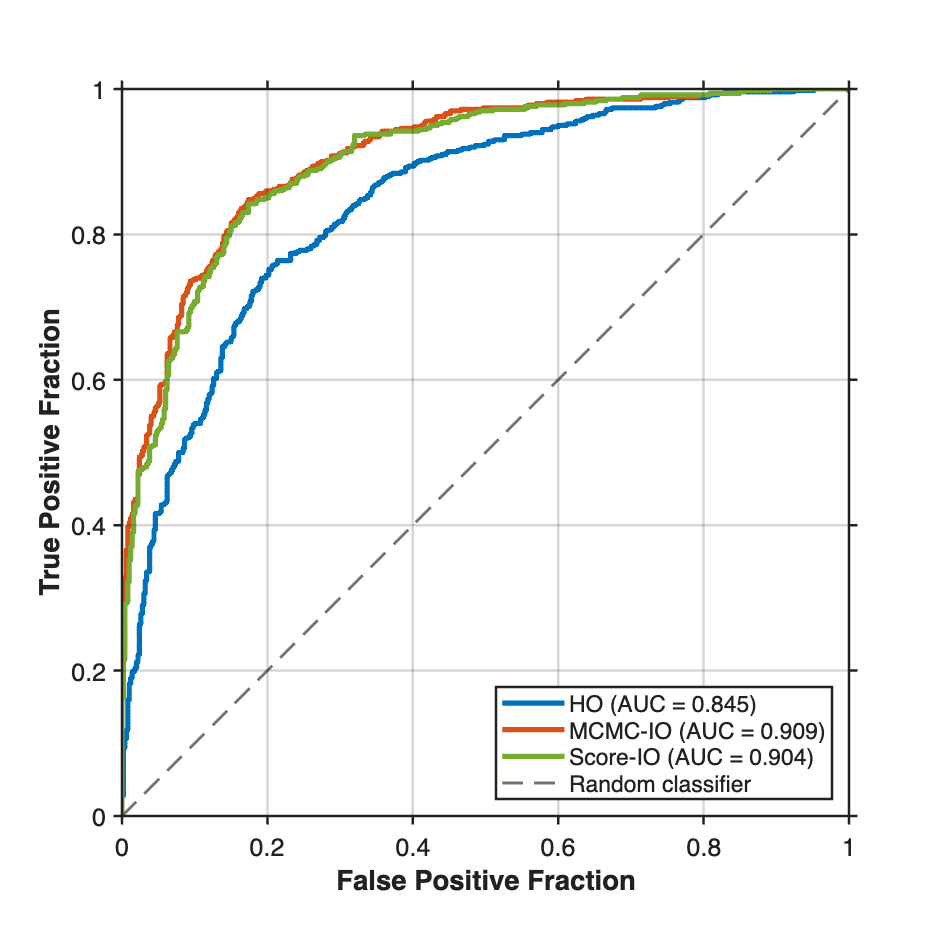}
         \vspace{-0.8cm}
         \caption{ROC curves of the SIO, MCMC-IO, and HO.}
     \end{subfigure}
     \vspace{0.2cm}
     \caption{Performance evaluation of the proposed score-based ideal observer (SIO). (a) The SIO performance rapidly converges as the number of integration points increases, and $K=5$ was used in subsequent experiments. (b) The proposed SIO closely approximates the MCMC-IO and substantially outperforms the HO.}
     \label{fig:AUC}
     \end{figure}

To investigate the numerical approximation of the proposed line-integral formulation in Eq. (\ref{eq:sio}), the SIO  performance was evaluated as a function of the number of integration points $K$. Figure~\ref{fig:AUC}(a) shows the AUC achieved by the proposed SIO for different values of $K$. The observer performance rapidly converges as $K$ increases and becomes nearly unchanged for $K\geq5$. Consequently, $K=5$ was employed in all subsequent experiments.

To evaluate the ability of the proposed SIO to approximate the IO, the MCMC-based ideal observer (MCMC-IO) was employed as a numerical reference for the IO performance. In addition, the Hotelling observer (HO) implemented using covariance matrix decomposition was included as a conventional linear-observer benchmark. Figure~\ref{fig:AUC}(b) compares the ROC curves of the proposed SIO, MCMC-IO, and HO. The ROC curve produced by the proposed SIO closely matches that of the MCMC-IO and substantially outperforms the HO, indicating that the proposed score-based formulation can accurately approximate the IO performance. Unlike MCMC-based approaches that require extensive posterior sampling for each test image or supervised learning approaches that may require task-specific retraining, the proposed method requires only a single DnCNN trained on signal-absent images and a small number of score evaluations along the signal interpolation path.

\section{Conclusion}

A novel score-based ideal observer (SIO) is proposed by expressing the Bayesian IO test statistic as the negative inner product between the known signal and an integrated score function. The proposed approach enables efficient IO approximation using a single denoising network trained only on signal-absent images, thereby eliminating the need for task-specific supervised retraining and computationally intensive posterior sampling during inference. Preliminary results demonstrate that the proposed SIO closely approximates Bayesian IO performance while substantially outperforming the Hotelling observer for the considered binary detection task involving a lumpy background model. Future work will investigate more realistic stochastic object models and inference tasks relevant to medical imaging applications.

\acknowledgments 
 
This work was supported by startup funds provided by the Wyant College of Optical Sciences and the Department of Radiology and Imaging Sciences at the University of Arizona. 

\bibliography{report} 
\bibliographystyle{spiebib} 

\end{document}